\documentclass[aps,pre,showpacs,showkeys,reprint,floatfix]{revtex4-1}

\usepackage{amssymb}
\usepackage{amsmath}
\usepackage{graphicx}
\usepackage{amsbsy}

\newcommand{\bq}{\begin{eqnarray}}
\newcommand{\eq}{\end{eqnarray}}
\newcommand{\bqn}{\begin{eqnarray*}}
\newcommand{\eqn}{\end{eqnarray*}}
\newcommand{\rr}{\mathbf{r}}

\newcommand{\nn}{{\bf n}}

\newcommand{\calh}{{\cal H}}
\newcommand{\calt}{{\cal T}}
\newcommand{\calv}{{\cal V}}

\begin{document}
\title{Effect of quantum dispersion on the radial distribution
function of a one-component sticky-hard-sphere fluid}

\author{Riccardo Fantoni}
\email{rfantoni@ts.infn.it}
\affiliation{Universit\`a di Trieste, Dipartimento di Fisica, strada
  Costiera 11, 34151 Grignano (Trieste), Italy}

\date{\today}

\begin{abstract}
In this short communication we present a possible scheme to study 
the radial distribution function of the quantum slightly polydisperse
Baxter sticky hard sphere liquid at finite temperature thorugh a
semi-analytical method devised by Chandler and Wolynes. 
\end{abstract}

\keywords{Quantum Sticky Hard Spheres, Polydispersity, Path Integral,
Reference Interaction Site Model}

\pacs{05.30.-d,61.20.-p,61.46.Df,82.70.Dd,83.80.Hj}

\maketitle

It is well known that a one-component classical Sticky-Hard-Sphere
(SHS) liquid \cite{Baxter68} is thermodynamically unstable
\cite{Stell91}. 

Nonetheless, when studied with Monte Carlo computer simulation the
fluid {\sl is} stable \cite{Miller03}. This is due to the
fact that a computer can only work with numbers with a finite number
of decimal figures. The computer arithmetics in fact differs from the
arithmetics of real number because the standard representation of
numbers must use a finite and fixed number of bits. So that the fluid
studied through the computer simulation will necessarily be
polydisperse (in size). And it has been proven that the polydisperse
SHS fluid is indeed thermodynamically stable \cite{Stell91}. 

It is then legitimate to pose the following questions: what would the
outcome for the radial distribution function of a quantum SHS fluid,
obeying to Boltzmann statistics (for the sake of simplicity),
calculated through the path integral Monte Carlo simulation, be? Can
one find a reasonable approximation for it, through other means?
The relevant parameters of the problem will be
the inverse temperature $\beta=1/K_BT$, the density $\rho$, the
spheres mass $m$ and diameter $\sigma$, and $\alpha$ the adhesion
coefficient. 

Aim of the note is to show how one may try to answer this
questions using an approach devised by Chandler and Wolynes
\cite{Chandler81} which relies on an {\sl isomorphism} between the
quantum statistical mechanics of a many body system and the classical
statistical mechanics of a particular {\sl 
polyatomic} fluid. Using the path integral formulation of quantum
statistical mechanics it can be shown (see appendix \ref{app:pa}) that
the canonical 
partition function of a system of $N$ quantum identical particles
of mass $m$ obeying to Boltzmann statistics and interacting through a
pair potential $v(r)$, at absolute temperature $T$, is approached in
the $P\to\infty$ limit by the classical partition function of $N$
indistinguishable ring {\sl molecules} made up of $P$ distinguishable
{\sl atoms}, at temperature $TP$, with a total potential energy
\bq \nonumber
&&V(R_0,\ldots,R_{P-1})=\\
&&\sum_{t=0}^{P-1}\left\{
\frac{|R_t-R_{t+1}|^2}{4\lambda\varepsilon^2} 
+\sum_{i<j}^Nv(|\rr_i^{(t)}-\rr_j^{(t)}|)\right\},
\eq
where $R_t\equiv(\rr_1^{(t)},\ldots,\rr_N^{(t)})$ are the positions of
the atoms at {\sl site} (imaginary thermal time slice) $t$ of the $N$
molecules, with 
$R_p=R_0$, and  
\bq
\lambda&=&\frac{\hbar^2}{2m},\\
\varepsilon&=&\frac{\beta}{P}.
\eq
This is known as the primitive action as explained in the appendix.

Note that for the SHS Baxter model \cite{Baxter68} one has 
\bq \label{baxter}
e^{-\beta v(r)}-1\xrightarrow{\mbox{sticky limit}}
-\theta(\sigma-r)+\sigma\alpha\delta(r-\sigma),
\eq
where the adhesion coefficient $\alpha\epsilon/\epsilon_0=1/12\tau$,
with $\epsilon_0$ a characteristic energy scale, is 
a {\sl monotonous function} of $\beta$. We can say that $\tau=\tau(\beta)$
is a monotonously increasing function of the absolute temperature $T$
representing a reduced temperature. The problem is then well set only
upon assigning the function $\tau(\beta)$.

The radial distribution function of the quantum system is then given
by 
\bq
g(r;\beta)=\lim_{P\to\infty}\frac{1}{P}\sum_{t=0}^{P-1}g_{0t}(r;\beta/P),
\eq
where $g_{tt^\prime}$ is the intermolecular site-site radial
distribution function of the isomorphic classical system.

The idea of Chandler and Wolynes is to use the Reference Interaction
Site Model (RISM) theory \cite{Hansen-McDonald} to determine the
$g_{0t}$ for $t=0,\ldots,P-1$ for a given $P$ ($P=2$ being the
simplest but less accurate approximation). That is, one needs to solve
the following integral equation subject to a given closure
\bq \nonumber
&&\mathbf{\hat{h}}(k;\varepsilon)=\\ \label{eq:RISM}
&&\mathbf{\hat{\pmb{\omega}}}(k;\varepsilon)
\mathbf{\hat{c}}(k;\varepsilon)[1-
\rho\mathbf{\hat{\pmb{\omega}}}(k;\varepsilon)
\mathbf{\hat{c}}(k;\varepsilon)]^{-1}
\mathbf{\hat{\pmb{\omega}}}(k;\varepsilon),
\eq
where $\mathbf{\hat{h}}(k;\varepsilon)$ and
$\mathbf{\hat{c}}(k;\varepsilon)$ are the matrices whose elements are
the Fourier transform of the intermolecular site-site total
correlation function
$h_{tt^\prime}(r;\varepsilon)=g_{tt^\prime}(r;\varepsilon)-1$ and
direct correlation function $c_{tt^\prime}(r;\varepsilon)$
respectively and the elements of
$\mathbf{\hat{\pmb{\omega}}}(k;\varepsilon)$ are the Fourier transform
of  
\bq
\omega_{tt^\prime}(\rr;\varepsilon)&=&\delta_{tt^\prime}\delta(\rr)+
(1-\delta_{tt^\prime})s_{tt^\prime}(r;\varepsilon),\\ \nonumber
&=&\left\{
\begin{array}{ll}
\delta(\rr) & t=t^\prime\\
s_{tt^\prime}(r;\epsilon) & t\neq t^\prime
\end{array}\right..
\eq  
where $s_{tt^\prime}(r;\varepsilon)$ are the intramolecular site-site radial
distribution functions of the isomorphic classical system, for which a
reasonable approximation is  
\bq \label{stt}
s_{tt^\prime}(r;\varepsilon)\approx \gamma_{tt^\prime}\,
e^{-\frac{r^2}{4\lambda|t-t^\prime|\varepsilon}}
y_{SHS}(r;\tau(|t-t'|\epsilon)),
\eq
where the normalization constant $\gamma_{tt^\prime}$ should be determined
from the condition
\bq
\int s_{tt^\prime}(r;\varepsilon)\, d\rr=1,
\eq
and $y_\text{SHS}(r;\tau)$ is the cavity radial distribution
function of a system of classical SHS of diameter $\sigma$, with
reduced temperature $\tau$ at a packing fraction
$\eta=\pi\rho\sigma^3/6$, $\rho=N/V$ being the density. That is
$y_\text{SHS}(r;\tau)=g_\text{SHS}(r;\tau)\exp[\tau v(r)]$ which is a
continuous function of $r$ even when the radial distribution function
of the SHS model, $g_\text{SHS}$, and/or $v$ are
discontinuous. 

In Eq. (\ref{stt}) the exponential factor stems from the
kinetic part of the action and again we used the functional dependence
of the adhesion coefficient $\tau$ on the inverse temperature
$|t-t'|\varepsilon$. 

\begin{figure}[htbp]
\begin{center}
\includegraphics[width=10cm]{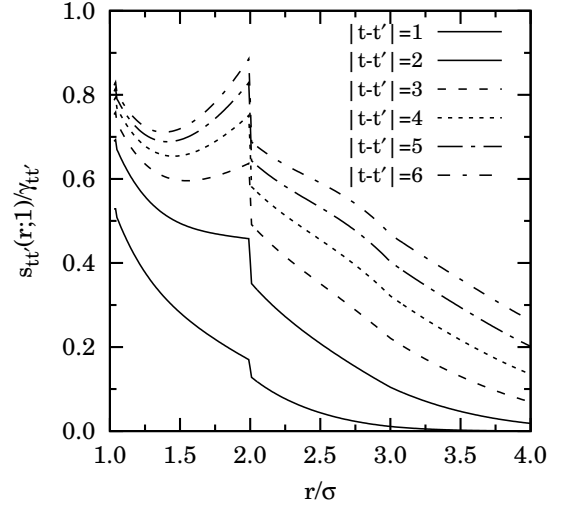}
\end{center}  
\caption{The intramolecular site-site radial distribution functions of
  the isomorphic classical system $s_{tt^\prime}(r;1)$ for
  $\lambda=1$, $\eta=0.32$, and $\tau=0.2$.}    
\label{fig:sr}
\end{figure}

Clearly we will have
$y_\text{SHS}(r;\tau)=g_\text{SHS}(r;\tau)$ for  
$r>\sigma$. The Laplace transform of $rg_\text{SHS}(r;\tau)$ in the
Percus-Yevick approximation for the SHS system is given by
\cite{Andres-Book}
\bq \label{gshslt}
\widehat{G}_\text{SHS}(s)&=&\int_0^\infty
dr\,e^{-sr}rg_\text{SHS}(r;\tau)\\ \nonumber
&=&\frac{e^{-s}}{s^2}\frac{\Lambda_0+\Lambda_1s+\Lambda_2s^2}
{1-12\eta[\varphi_2(s)\Lambda_0+\varphi_1(s)\Lambda_1+\varphi_0(s)\Lambda_2]},
\eq
where,
\bq
\varphi_k(x)=x^{-(k+1)}\left(\sum_{l=0}^k\frac{(-x)^l}{l!}-e^{-x}\right),
\eq
and
\bq
\Lambda_0&=&\frac{1+2\eta}{(1-\eta)^2}-\frac{12\eta}{1-\eta}\Lambda_2,\\
\Lambda_1&=&\frac{1+\eta/2}{(1-\eta)^2}-\frac{6\eta}{1-\eta}\Lambda_2.\\
\Lambda_2&=&\frac{1-(1-\tau^{-1})\eta-w}{2\tau^{-1}(1-\eta)\eta},\\ \nonumber
w&=&\sqrt{(1-\eta)\left[1-\eta\left(1-2\tau^{-1}+
\frac{\tau^{-2}}{3}\right)\right]+\frac{\tau^{-2}}{2}\eta^2},
\eq
In Fig. \ref{fig:sr} we show the intramolecular site-site radial
distribution functions of the isomorphic classical system assuming an
adhesion coefficient independent from temperature. 

For the closure one may use the modified Mean Spherical Approximation
(mMSA) \cite{Gazzillo04}
\bq
c_{tt^\prime}(r)=f_{tt^\prime}(r)=e^{-\beta v_{tt^\prime}(r)}-1,
~~~r>d_{tt^\prime},
\eq
where
\bq
v_{tt^\prime}(r)=\left\{
\begin{array}{ll}
v(r) & t=t^\prime\\
0    & t\neq t^\prime
\end{array}\right.,
\eq
and 
\bq \label{dtt}
d_{tt^\prime}=\left\{
\begin{array}{ll}
\sigma & t=t^\prime\\
0      & t\neq t^\prime
\end{array}\right.,
\eq
Here we are neglecting the fact that the size of a path (or polymer),
its thermal wavelength, is $\Lambda_\beta=\sqrt{\beta\hbar^2/m}$.
\footnote{One may take into account of the size of the path by taking
  for example 
$d_{tt^\prime}
=\{\sigma ~~\mbox{for}~~ t=t^\prime,~~
\{ 0 ~~\mbox{for}~~ \sigma<\Lambda_\beta , \sigma-\Lambda_\beta
~~\mbox{for}~~ \sigma>\Lambda_\beta\} 
~~\mbox{for}~~ t=t^\prime\pm 1,~~
0  ~~\mbox{otherwise}\}$.}
Combined with the exact relation valid for $r\le d_{tt^\prime}$
\bq
h_{tt^\prime}(r)=\left\{
\begin{array}{ll}
\frac{\sigma}{12\tau}y_{tt}(d_{tt})\delta(r-d_{tt})-1 & t=t^\prime\\
-1                                                 & t\neq t^\prime
\end{array}\right.,
\eq
where $y_{tt}(d_{tt})$ are the intermolecular site-site cavity
functions at contact which in the mMSA are \cite{Gazzillo04}
\bq
y_{tt}(d_{tt})=1,~~~t=0,\ldots,P-1.
\eq
Then, for the closure, we will have
\bq
c_{tt^\prime}(r)&=&0,~~~r>d_{tt^\prime},\\
h_{tt^\prime}(r)&=&\left\{
\begin{array}{ll}
\frac{\sigma}{12\tau}\delta(r-\sigma)-1 & t=t^\prime\\
-1                                      & t\neq t^\prime
\end{array}\right.,~~~r\le d_{tt^\prime}.
\eq

The RISM integral equation (\ref{eq:RISM}) can be rewritten as the
following Ornstein-Zernike-like relation,
\bq \nonumber
&&\mathbf{\hat{h}}(k;\varepsilon)=\\ \label{eq:RISM1}
&&\mathbf{\hat{\pmb{\omega}}}(k;\varepsilon)
\mathbf{\hat{c}}(k;\varepsilon)\mathbf{\hat{\pmb{\omega}}}(k;\varepsilon)+
\rho\mathbf{\hat{\pmb{\omega}}}(k;\varepsilon)
\mathbf{\hat{c}}(k;\varepsilon)
\mathbf{\hat{h}}(k;\varepsilon).
\eq
 
The main obstacle in solving this integral equation reside in the fact
that the intramolecular site-site radial distribution function of the
isomorphic classical system, $s_{tt'}(r;\epsilon)$, is known only
numerically through Laplace inversion of Eq. (\ref{gshslt}) obtained
for example using the algorithm of Abate and Whitt \cite{Abate1992}.

The uncontrolled approximations in this treatment reside in:
(i) Eq. (\ref{stt}), where we have approximated the full equilibrium
distribution function for $P$ {\sl cavities} forming a molecule with
the cavity pair distribution function of the SHS classical fluid (this
approximation becomes worse and worse as $P$ decreases). Since the
primitive approximation error goes like $\lambda\varepsilon^2$
\cite{Ceperley1995} it is reasonable to expect that a good enough
approximation would require $\lambda\beta^2/P^2\sim 0.01$. Of course
one reasonably expects that solving RISM equations numerically becomes
rapidly a difficult task (including non-convergence problems) as P
increases; (ii) Eq. (\ref{dtt}), where we are neglecting the thermal
wavelength of a polymer.

To our knowledge the quantum slightly polydisperse Baxter sticky hard
spheres liquid has never been studied before neither through computer
simulations of the one-component system nor through other means.
To asses the existence of thermodynamic and structural properties of 
such a physical model from a rigorous mathematical point of view
seems to be a quite formidable task. In this respect the theory of
path integrals should probably be the place where to start to look
at. It is infact out of doubt that at any finite $P$ the classical
isomorphic system is thermodynamically ($N\to\infty$ at constant
$\rho$) well defined, but understanding the effect of the 
slightly polydisperse adhesion (the last term in Eq. (\ref{baxter}))
in the $P\to\infty$ (Feynman-Kac-)limit does not seem so easy.
There are three different limits we have to deal with: (i) the sticky
limit, (ii) the path integral limit, and (iii) the thermodynamic
limit. While it is quite customary to take the thermodynamic limit in
the end, the order of the first two limits should be immaterial. 
Moreover we expect the path integral solution to dependent crucially
on the choice of the function $\tau(T)$. 

We plan to adopt the present scheme to obtain semi-analytical
quantitative results for the radial distribution function of the
extension to the quantum regime of some of the classical fluids
studied in Refs. \cite{Fantoni05a,Fantoni05b,Fantoni06a,Fantoni06b,Fantoni06c,Fantoni07,Fantoni08a,Fantoni09b,Fantoni11a,Fantoni12a,Fantoni13h,Fantoni14b,Fantoni15a,Fantoni15c,Fantoni16c},
in the near future.  

\appendix
\section{The primitive action}
\label{app:pa}
In this appendix we give a brief review of the derivation of the
primitive approximation given in Ref. \cite{Ceperley1995}.
Suppose the Hamiltonian is split into two pieces $\calh=\calt+\calv$,
where $\calt$ and $\calv$ are the kinetic and potential
operators. Recall the exact Baker-Campbell-Hausdorff formula to
expand $\exp(-\varepsilon\calh)$ into the product
$\exp(-\varepsilon\calt)\exp(-\varepsilon\calv)$. As $\varepsilon\to 0$ the commutator
terms which are of order higher than $\varepsilon^2$ become smaller than the
other terms and thus can be neglected. This is known as the {\sl
  primitive approximation}
\bq
e^{-\varepsilon(\calt+\calv)}\approx e^{-\varepsilon\calt}e^{-\varepsilon\calv}.
\eq
hence we can approximate the exact density matrix by product of the
density matrices for $\calt$ and $\calv$ alone. One might worry that
this would lead to an error as $P\to\infty$, with small errors
building up to a finite error. According to the Trotter
\cite{Trotter1959} formula, one does not have to worry
\bq
e^{-\beta(\calt+\calv)}=\lim_{P\to\infty}\left[e^{-\varepsilon\calt}e^{-\varepsilon\calv}\right]^P.
\eq
The Trotter formula holds if the three operators $\calt$, $\calv$, and
$\calt+\calv$ are self-adjoint and make sense separately, for example,
if their spectrum is bounded below. \cite{Simon1979} This is the case
for the Hamiltonian describing SHS.

Let us now write the primitive approximation in position space
$R=(\rr_1,\rr_2,\ldots,\rr_N)$ with $\rr_i$ the coordinate of the
$i$th particle, 
\bq
\rho(R_0,R_2;\varepsilon)\approx\int dR_1\langle
R_0|e^{-\varepsilon\calt}|R_1\rangle\langle R_1|e^{-\varepsilon\calv}|R_2\rangle,
\eq
and evaluate the kinetic and potential density matrices. Since the
potential operator is diagonal in the position representation, its
matrix elements are trivial
\bq \label{primitive-v}
\langle R_1|e^{-\varepsilon\calv}|R_2\rangle=e^{-\varepsilon V(R_1)}\delta(R_2-R_1).
\eq

The kinetic matrix can be evaluated using the eigenfunction expansion
of $\calt$. Consider, for example, the case of distinguishable
particles in a cube of side $L$ with periodic boundary
conditions. Then the exact eigenfunctions and eigenvalues of $\calt$
are $L^{-3N/2}e^{iK_\nn R}$ and $\lambda K_\nn^2$, with
$K_\nn=2\pi\nn/L$ and $\nn$ a $3N$-dimensional integer vector. We are
using here dimensional units. Then 
\bq \label{primitive-t-1}
\langle R_0|e^{-\varepsilon\calt}|R_1\rangle&=&\sum_\nn L^{-3N}e^{-\varepsilon\lambda
  K_\nn^2}e^{-iK_\nn(R_0-R_1)}\\ \label{primitive-t-2}
&=&(4\pi\lambda\varepsilon)^{-3N/2}
\exp\left[-\frac{(R_0-R_1)^2}{4\lambda\varepsilon}\right],
\eq
where $\lambda=\hbar^2/2m$. Eq. (\ref{primitive-t-2}) is obtained by
approximating the sum by an integral. This is appropriate only if the
thermal wavelength of one step is much less than the size of the box,
$\lambda\varepsilon\ll L^2$. In some special situations this condition could
be violated, in which case one should use Eq. (\ref{primitive-t-1}) or
add periodic ``images'' to Eq. (\ref{primitive-t-2}). The exact
kinetic density matrix in periodic boundary conditions is a theta
function, $\prod_{i=1}^{3N}\theta_3(z_i,q)$, where
$z_i=\pi(R_0^i-R_1^i)/L$, $R^i$ is the $i$th component of the $3N$
dimensional vector $R$, and $q=e^{-\lambda\varepsilon(2\pi/L)^2}$ (see chapter
16 of Ref. \cite{Abramowitz}). Errors from ignoring the boundary
conditions are $O(q)$, exponentially small at large $P$.  

A {\sl link} $m$ is a pair of time slices $(R_{m-1},R_m)$ separated by
a {\sl time step} $\varepsilon=\beta/P$. The {\sl action} $S^m$ of a link is
defined as minus the logarithm of the exact density matrix. Then the
exact path-integral expression becomes
\bq
\rho(R_0,R_P;\beta)=\int dR_1\ldots dR_{P-1}\,
\exp\left[-\sum_{m=1}^P S^m\right],
\eq 
It is convenient to separate out the {\sl kinetic action} from the
rest of the action. The exact kinetic action for link $m$ will be
denoted $K^m$ 
\bq \label{eq:primitive-k}
K^m=\frac{3N}{2}\ln(4\pi\lambda\varepsilon)+\frac{(R_{m-1}-R_m)^2}{4\lambda\varepsilon},
\eq
The {\sl inter-action} is then defined as what is left
\bq
U^m=U(R_{m-1},R_m;\varepsilon)=S^m-K^m.
\eq
In the primitive approximation the inter-action is 
\bq \label{eq:primitive-u}
U^m_1=\frac{\varepsilon}{2}[V(R_{m-1})+V(R_m)],
\eq
where we have symmetrized $U^m_1$ with respect to $R_{m-1}$ and $R_m$,
since one knows that the exact density matrix is symmetric and thus
the symmetrized form is more accurate. 

A capital letter $U$ refers to the total link inter-action. One should
not think of the exact $U$ as being strictly the potential
action. That is true for the primitive action but, in general, is only
correct in the small-$\varepsilon$ limit. The exact $U$ also contains kinetic
contributions of higher order in $\varepsilon$. If a subscript is present on
the inter-action, it indicates the order of approximation; the
primitive approximation is only correct to order $\varepsilon$. No subscript
implies the exact inter-action.

The {\sl residual energy} of an approximate density matrix is defined
as 
\bq \label{residual-energy}
E_A(R,R';t)=\frac{1}{\rho_A(R,R';t)}\left[\calh+
\frac{\partial}{\partial t}\right]\rho_A(R,R';t).
\eq
The residual energy for an exact density matrix vanishes; it is a
local measure of the error of an approximate density matrix. The
Hamiltonian $\calh$ is a function of $R$; thus the residual energy is
not symmetric in $R$ and $R'$. 

It is useful to write the residual energy as a function of the
inter-action. We find
\bq \nonumber
E_A(R,R';t)&=&V(R)-\frac{\partial U_A}{\partial t}-
\frac{(R-R')\cdot \nabla U_A}{t}+\\ \label{residual-energy-b}
&&\lambda\nabla^2U_A-
\lambda\left(\nabla U_A\right)^2.
\eq
The terms on the right hand side are ordered in powers of $\varepsilon$,
keeping in mind that $U(R)$ is of order $\varepsilon$, and $|R-R'|$ is of
order $\varepsilon^{1/2}$. One obtains the primitive action by setting the
residual energy to zero and dropping the last three terms on the right
hand side. 

The residual energy of the primitive approximation is
\bq \nonumber
E_1(R,R';t)&=&\frac{1}{2}\left[V(R)-V(R')\right]-
\frac{1}{2}(R-R')\cdot\nabla V+\\
&&\frac{\lambda t}{2}\nabla^2V-
\frac{\lambda t^2}{4}\left(\nabla V\right)^2.
\eq
With a leading error of $\sim \lambda\varepsilon^2$.

\begin{acknowledgments}
I would like to acknowledge friutful discussion with Prof. Domenico
Gazzillo during my stay at the Chemical Physics department of the
University Ca' Foscari of Venice.
\end{acknowledgments} 
%

\end{document}